\documentclass[dvipdfmx]{pasj01}
\Received{2015 Aug 19}
\Accepted{2015 Nov 5}
\Published{$\langle$publication date$\rangle$}
\newcommand{\feoh}{\textrm{[Fe/H]}}

\newcommand{\msun}{\ensuremath{M_{\odot}}}

\newcommand{\teff}{\ensuremath{T_{\textrm {eff}}}}
\newcommand{\pow}[2]{\ensuremath{{#1} \times 10^{#2}}}
\newcommand{\hyp}{\ensuremath{\, \mathchar`- \,}}
\newcommand{\menv}{\ensuremath{M_{\rm env}}}
\newcommand{\mcore}{\ensuremath{M_{\rm c}}}

\begin{document}

\title{On the Brightness of Surviving Companions in Type Ia Supernova Remnants}

\author{Kazuhiro Noda\altaffilmark{1,2}, Takuma Suda\altaffilmark{1}, and Toshikazu Shigeyama\altaffilmark{1}}
\altaffiltext{1}{Research Center for Early Universe, Graduate School of Science,The University of Tokyo, 7-3-1 Hongo, Bunkyo-ku, Tokyo 113-0033, Japan}
\altaffiltext{2}{Department of Astronomy, Graduate School of Science,The University of Tokyo, 7-3-1 Hongo, Bunkyo-ku, Tokyo 113-0033, Japan}
\email{noda@resceu.s.u-tokyo.ac.jp}


\maketitle

\begin{abstract}
The progenitor systems for type Ia supernovae are still controversial. One of the methods to test the proposed scenario for the progenitor systems is to identify companions that are supposed to survive according to the so-called single degenerate scenario. These companions might be affected by supernova ejecta. We present several numerical simulations of surviving red-giant companions  whose envelopes were stripped and heated. We find that red-giants with less-massive helium cores ($\lesssim0.30~\msun$)  can be  so faint after the supernovae  that we cannot detect them. In addition,  we apply  the results to  the case of SNR 0509-67.5, and   put constraints on the helium core mass, envelope stripping, and energy injection  under the single degenerate scenario for type Ia supernovae. 
\end{abstract}

\begin{table*}[ht]
%\rotate
%\tablewidth{20 cm}
\tbl{ Characteristics of models prior to the envelope stripping. \label{tab:initmodel}}{
%\startdata
%\begin{center}
\begin{tabular}{l l l l l l l l} \hline \hline
$M_{\rm c}(M_\odot)$ & $t_{\rm rlof}$ (Gyr) & $\log R_{\rm c}(R_\odot)$ & $\log \rho_{\rm c,s}({\rm g/cm^3})$ & $\log R(R_\odot)$ & $\log L(L_\odot)$ & $\log E_{\rm bind,env}({\rm erg})$ & $\log H_{\rm p(c,s)}({\rm cm})$ \\
\hline
0.101 & 0.623 & -1.60 & 3.56 & 0.35 & 0.50 & 48.27 & 8.50 \\
0.132 & 0.744 & -1.59 & 3.48 & 0.43 & 0.62 & 48.16 & 8.50 \\
0.162 & 0.878 & -1.59 & 3.35 & 0.54 & 0.83 & 48.03 & 8.45 \\
0.190 & 0.987 & -1.59 & 3.23 & 0.69 & 1.10 & 47.89 & 8.42 \\ 
0.220 & 1.061 & -1.60 & 3.12 & 0.88 & 1.43 & 47.74 & 8.34 \\
0.250 & 1.103 & -1.61 & 3.07 & 0.99 & 1.61 & 47.65 & 8.28 \\
0.285 & 1.138 & -1.61 & 2.84 & 1.25 & 2.03 & 47.45 & 8.29 \\ 
0.349 & 1.159 & -1.56 & 2.07 & 1.65 & 2.63 & 47.06 & 8.29 \\
0.400 & 1.165 & -1.56 & 1.87 & 1.91 & 3.01 & 46.91 & 8.27 \\
\hline
\end{tabular}}
%\end{center}

\label{tab:initmodel} 
\end{table*}

\section{Introduction}

Type Ia supernovae (SNe Ia) are luminous astrophysical events and are thought to be explosions of white dwarfs (WDs) in binary systems. A strong correlation between the peak luminosities and the timescales of the subsequent declines \citep{Phillips1993} have made SNe Ia useful tools to determine the cosmological constants (e.g., \authorcite{Riess1998} \yearcite{Riess1998}; \authorcite{Schmidt1998} \yearcite{Schmidt1998}; \authorcite{Perlmutter1999} \yearcite{Perlmutter1999}). In spite of their importance, the progenitor systems of SNe Ia are not identified yet and there are two competing scenarios (see \authorcite{Maoz2012} \yearcite{Maoz2012}, for a recent review). One of them is called the single degenerate (SD) scenario \citep{Whelan1973}, which involves a carbon-oxygen white dwarf and a non-degenerate companion. In this scenario, the WD increases its mass by accreting gas from the non-degenerate companion and explodes when the mass reaches the Chandresekhar limit. The other is the double degenerate (DD) scenario \citep{Iben1984,Webbink1984} in which two CO WDs merge and form an object with a mass exceeding the Chandrasekhar limit that eventually leads to a thermonuclear explosion.

Several observational trials have been performed to search for signs to identify the progenitor system. Actually, the optical spectra of an SN Ia PTF 11kx implying multiple components of circumstellar material  are naturally explained by a symbiotic nova system. Thus this particular SN Ia is likely to originate from a binary system with a red-giant companion and supports the SD scenario \citep{Dilday2012}. \citet{Ruiz2004} performed a direct search for a surviving companion in Tycho's supernova remnant (SNR) and identified a type G0-2 star as a candidate for the surviving companion, however this is still controversial (e.g., \authorcite{Fuhrmann2005} \yearcite{Fuhrmann2005}; \authorcite{Ihara2007} \yearcite{Ihara2007}; \authorcite{Gonzalez2009} \yearcite{Gonzalez2009}; \authorcite{Kerzendorf2009} \yearcite{Kerzendorf2009}). On the other hand, \citet{Schaefer2012} examined an image of SNR 0509-67.5 taken by the Hubble Space Telescope and put an upper limit of $M_{\rm V}>+8.4$ to the brightness of the surviving companion star, which apparently excludes the existence of any companion stars predicted by the SD scenario and favors the DD scenario. On the other hand, it is argued that the SN responsible for SNR 0509-67.5 is classified as an over-luminous SN and originates from the SD channel \citep{Fisher2015}.

An SN Ia is expected to significantly affect the  companion star under the SD scenario.  For example, the companion star might lose its  envelope due to an impact of the SN ejecta. \citet{Wheeler1975} analytically calculated the amount of the stripped mass and found that main sequence stars are only slightly stripped, while red-giants lose their entire envelopes. Two dimensional numerical simulations showed that main sequence and sub-giant companions lose 15\% of their mass, while a red-giant companion loses 96\%-98\% of the envelope as a result of the impact of SN ejecta \citep{Marietta2000}. \citet{Podsiadlowski2003} simulated the evolution of a surviving sub-giant companion and concluded that the appearance of the companion is strongly affected by envelope stripping and energy injection by the impact of SN ejecta. So far, very few studies have concentrated on the evolution of surviving companions.

In the context of intense mass loss processes in a close binary system, a helium white dwarf can be formed by the Roche lobe overflow \citep{Kippenhahn1968, Webbink1975, Iben1986}. In their calculations for a red-giant with a $\sim 0.3{\,M_\odot}$ helium core, the hydrogen burning rate suddenly decreases when the mass of the hydrogen rich envelope is reduced to $\sim 2\times10^{-3}{\,M_\odot}$. More detailed calculations revealed the evolution of resultant helium white dwarfs with several different timings of the mass loss episode \citep{Althaus1997}. These studies were concerned with the long term evolution of helium cores  after the mass loss episodes over periods of the order of million years or longer.  On the other hand,  targets of searches for surviving companion stars are necessarily young SNRs with ages of several hundred years or less because we need to identify the types of the SNe by their spectra. For example, Tycho's SNR and SNR 0509-67.5, both of which are $\sim$500 yr old, were classified as type Ia by analyzing spectra of the light echoes \citep{Krause2008, Rest2008}. 

Thus the purpose of this paper is to investigate the evolution of surviving companions in the SD scenario after SN explosions on the timescale of $\sim$1,000 yr. We focus on binary systems with red-giants because the SN explosions are expected to strip significant fractions of the envelopes, and should affect the subsequent evolution. We apply the results to the case of SNR 0509-67.5 and explore the possibility of finding a surviving companion.

\section{Models of non-degenerate companions}

We follow the evolution of companions with a stellar evolution code \citep{Suda2010, Iben1992}. Input physics are the same as in \citet{Suda2011}. We adopt the nuclear reaction rates of NACRE \citep{Angulo1999} for proton-, electron- and $\alpha$-capture reactions involving nine nuclides. We employ the radiative opacity tables based on OPAL \citep{Iglesias1996} and \citet{Alexander1994} with the same procedure for interpolation as in \citet{Suda2010}.
Conductive opacities are computed using the fitting formulae of \citet{Itoh2008}.
%Conductive opacities were computed using the fitting formulae of \citet{Itoh1983} with the correction of the electro-static force \citep{Itoh2004}.
Neutrino energy losses are taken from \citet{Itoh1996}.
Mass loss is ignored during the ascent on the red-giant branch, i.e., the total mass $M_{\ast}$ of a star does not change for the fiducial model.
The convective mixing is treated by the mixing length theory described in \citet{Iben2013}.
We use the Schwarzschild criterion to determine the convective boundaries with the mixing length parameter of $\alpha = 1.5$.
The initial mass $M_{\ast, 0}$ and metallicity $Z$ of model stars are fixed to be $1 \msun$ and $0.01$, respectively throughout this study.
We chose the metallicity to apply our models to type Ia SNe  in the Large Magellanic Cloud where the typical value of the metallicity is $\feoh = -0.40$ for red-giants \citep{Cole2005}.
We follow the evolution of stars from the zero-age main sequence  to the tip of the red-giant branch (RGB).
%\citep{Cole2005}.

To simulate the envelope stripping by SN ejecta at various evolutionary stages of the companion star, we extract red-giant models with core masses of
$\mcore=$0.101, 0.132, 0.162, 0.190, 0.220, 0.250, 0.285, 0.349, and $0.400~\msun$.
Here we have defined the core mass $\mcore$ as the value of the enclosed mass of the shell  below which the mass fraction of hydrogen equals to zero.

The properties of these nine pre-supernova companions are shown in Table~\ref{tab:initmodel}.
Each column denotes the core mass $\mcore$, the time $t_{\rm rlof}$ since the formation of an isothermal core , the core radius $R_{\rm c}$, the density at the surface of the core $\rho_{\rm c,s}$, the stellar radius $R$, the stellar luminosity $L$, the binding energy $E_{\rm bind,env}$ of the envelope, and the pressure scale height $H_{\rm p(c,s)}=\left|{\rm d}r/{\rm d}\log p\right|$ at the surface of the core.
Our models assume that a red-giant with a core mass of $\mcore$ fills the Roche Lobe and starts to supply mass to the white dwarf at $t_{\rm rlof}$ and  that the white dwarf explodes in a few tens of Myr meanwhile the core of the red-giant does not significantly grow.
It is to be noted from Table 1 that  the timescale $t_{\rm rlof}$ is not so sensitive to the stellar mass as the the main sequence lifetime.

We remove the outer layers of companions on the timescale of $\sim6\times10^6$ sec  or less to mimic the envelope stripping by SN ejecta \citep{Marietta2000}.
We parameterize the degree of  the envelope stripping by the residual envelope mass $M_{\rm env}$, which is defined as $\menv=M_{\ast}-\mcore$  at the end of the envelope stripping .
We regard this envelope mass as a free parameter and make $\sim10 \hyp 20$ models for each core mass with various envelope masses in the range of $M_{\rm env}=5\times10^{-1}~\msun \hyp 2\times10^{-6}~\msun$. 

\begin{figure}[t]
 % \epsscale{0.9}
 \begin{center}
  \includegraphics[width=0.45\textwidth]{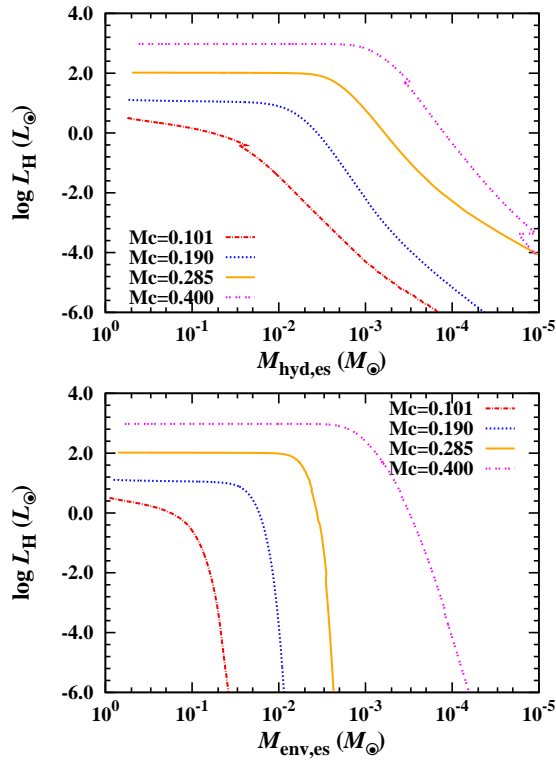}
  \end{center}
  \caption{Temporal change of hydrogen burning luminosity during the envelope stripping as functions of hydrogen mass in the envelope (top) and the envelope mass (bottom). }
  \label{fig:nuc}
\end{figure}

\begin{figure}[t]
 % \epsscale{0.7}
 \begin{center}
  \includegraphics[width=0.45\textwidth]{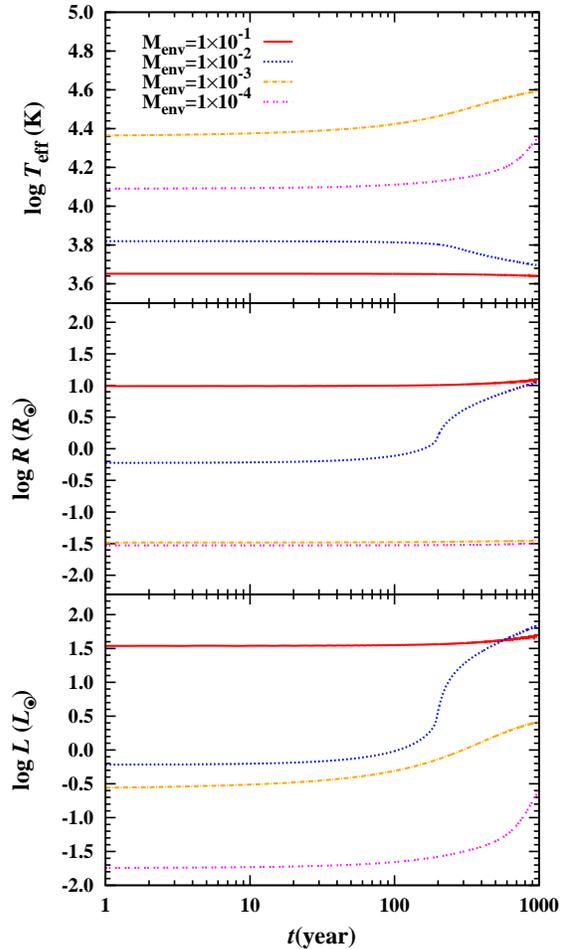}
  \end{center}
  \caption{Time evolution of the effective temperature (top), radius (middle), and luminosity (bottom) of companions with $\mcore=0.285~\msun$ after the envelope stripping. The energy injected into the envelope is not taken into account. Each line corresponds to a different residual envelope mass as indicated in the top panel.}
  \label{fig:evo285}
\end{figure}

In addition, we consider the extra energy injected into the envelopes of red-giants by the impacts of SN ejecta for some models.
This may be represented by an additional internal energy in the remaining envelope immediately after the envelope stripping.
%, with the arbitrary amount of energy in each timestep.}
 To see the effect on the envelope inflation by this procedure, we chose models with the following combinations of the helium core mass and the envelope mass; $\mcore=0.132$ and 0.190 $\msun$ and $\menv=1\times10^{-2},1\times10^{-3},\,{\rm and}\,1\times10^{-4}~\msun$.
We inject the specified energy within 1 yr at a constant rate.
 Since the detailed processes of energy injection into the envelope is poorly understood, we simply  distribute the same amount of energy per unit mass throughout the envelope.
The total  injected energy is approximately  set to a quarter or one third of the total gravitational binding energy of the envelope.

We follow the evolution of  models of companion stars for $\sim$1,000 yr
 after explosion to  safely cover the age of  SNR 0509-67.5 ($\sim400$ yr according to \citet{rest2005}).

\begin{figure*}[ht]
%  \epsscale{1.0}
\begin{center}
  \includegraphics[width=0.9\textwidth]{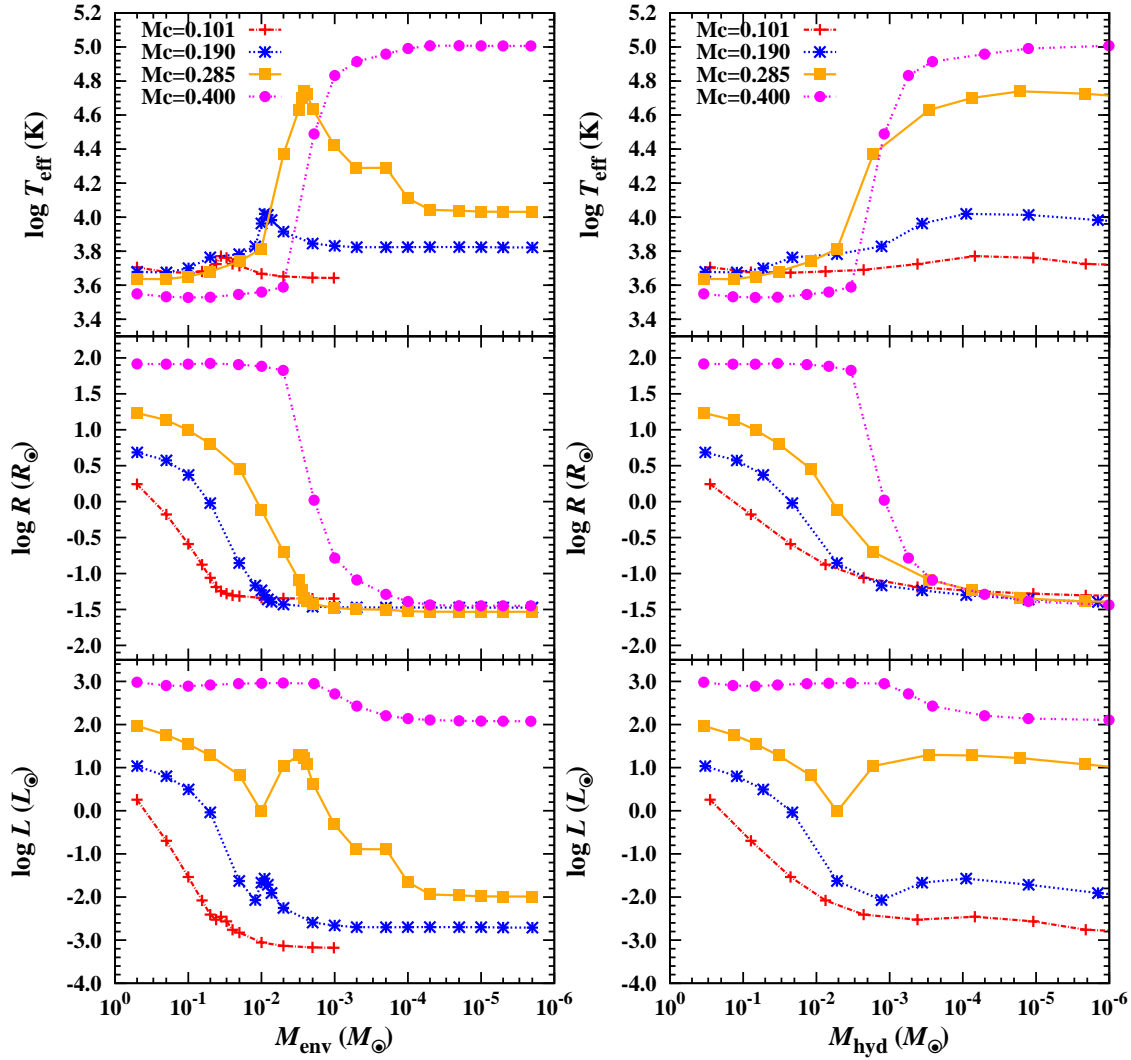}
  \end{center}
  \caption{Effective temperature  (top), radius  (middle) , and luminosity (bottom) of companions as functions of the envelope mass (left) and the hydrogen mass in the envelope (right)  at 100 yr after the explosion (or the envelope stripping).  The energy injected into the envelope is not taken into account.   The points for a model with the same core mass are connected by a  line as indicated in the top panels.}
  \label{fig:100yr}
\end{figure*} 
\begin{figure}[ht]
 \includegraphics[width=0.45\textwidth]{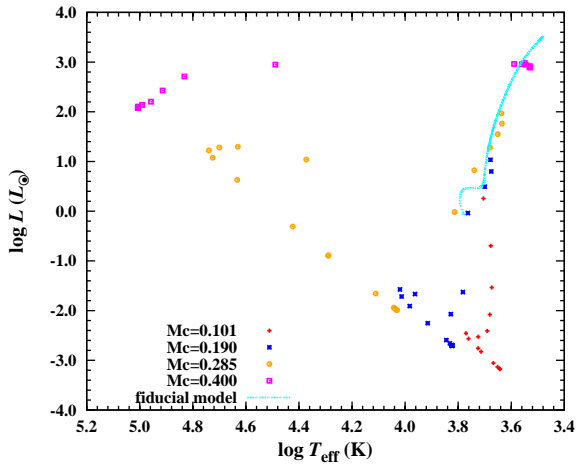}
  \caption{ Locations of companions on the H-R diagram  at 100 yr after the explosion (or the envelope stripping).  The energy injected into the envelope is not taken into account. The symbols denote the helium core mass as indicated in the panel. Individual points with the same symbol represent  different values of $\menv$ lined up along the sequence from the red-giant to white dwarf branch. The cyan dotted line shows the evolutionary track of a star with an initial mass of $1 \msun$ and $Z=0.01$ from the zero-age main sequence to the tip of the red-giant branch.}
  \label{fig:100HR}
\end{figure}

\section{Results}
\subsection{Models without  energy injections}

 We begin with the evolution of companion stars  after the envelope stripping without energy injection.
The remaining envelope is assumed to be unaffected by SN blast waves.
This assumption may give a lower limit to the  resultant luminosities  for given $\mcore$ and $\menv$. 

Figure \ref{fig:nuc} shows the temporal change  in the energy generation rates during the envelope stripping as functions of the envelope mass for red-giants with $\mcore=$0.101, 0.190, 0.285, and 0.400 $\msun$.
Since the envelope is stripped by the impact of SN ejecta on a much shorter timescale than the diffusion timescale or the thermal adjustment timescale, the envelope does not expand and  the hydrogen flash is not triggered in the thin envelope. These features are in contrast with what was observed in simulations of red-giants with  stripping envelopes on much longer timescales \citep{Kippenhahn1968,Iben1986,Driebe1999}.
The  hydrogen burning is extinguished when $M_{\rm env, es} \lesssim \pow{7}{-3} ~\msun$  for   a $\mcore = 0.285~\msun$, which is consistent with  results for the model  with $\mcore\sim 0.3~\msun$ of \citet{Iben1986}.
Here we have defined the envelope mass $M_{\rm env, es}$ as $M_{\ast}-\mcore$ during the envelope stripping.
 The  hydrogen burning rate increases with  increasing core mass for a given envelope mass $M_{\rm env, es}$ or the corresponding mass $M_{\rm hyd, es}$ of hydrogen in the envelope.
This is because the lower temperature of the burning shell for a smaller core mass drastically reduces the nuclear energy generation rate by the CNO cycles.
Obviously, the  subsequent  evolution is strongly affected by this decline of the energy generation rate.

The evolution after the envelope stripping is computed with usual setups with mesh rezoning.
The companion model with $M_{\rm c}=0.285~\msun$ shows an abrupt drop of the energy generation rate at $M_{\rm env, es}=2 \hyp 3\times10^{-3}~\msun$ (Fig. \ref{fig:nuc}).
Figure \ref{fig:evo285} shows the  temporal evolution of  the effective temperature $T_{\rm eff}$, radius $R$, and luminosity $L$ for $M_{\rm c}=0.285 ~\msun$ with different residual envelope masses of $\menv = \pow{1}{-1}$, $\pow{1}{-2}$, $\pow{1}{-3}$, and $\pow{1}{-4}~ \msun$.
As seen from the  figure, the structure of the envelope is bifurcated by a critical mass of the envelope  between $10^{-2}$ and $10^{-3} ~\msun$.
Models with $\menv > 10^{-2} \msun$  continue the hydrogen burning. These stars  become faint immediately after the envelope stripping but eventually return to the RGB in $\sim 1,000$ yr, while models with $\menv$  below the critical mass do not recover the hydrogen burning and evolve to white dwarfs. Their luminosities at $\sim 1,000$ yr are sensitive to $\menv$ and decrease with  decreasing $M_{\rm env}$ because of decreasing amount of nuclear fuels.
 The critical envelope mass  is estimated to be $ 2 \hyp \pow{3}{-3} ~\msun$ for this particular core mass.
Therefore, we need to investigate possible outcomes from the envelope stripping to assess the detectability of He-WD companions around type Ia SNe.

%\subsubsection{Appearences of the companion at 100 years after explosion}

Figure \ref{fig:100yr}  summarizes $T_{\rm eff}$, $R$, and $L$ as functions of $M_{\rm env}$ or $M_{\rm hyd}$  for the models presented in Figure \ref{fig:nuc} at $100$ yr after the envelope stripping. Here $M_{\rm hyd}$ denotes the total mass of remaining hydrogen  in the star.

These models have various thermal timescales spanning a few orders of magnitude depending on the residual mass of the envelope. Some models evolve in 100 yr after  the stripping while some models do not significantly change their appearance for 1,000 yr.  For example,  the luminosity still increases for $\menv \sim 10^{-2} \msun$ as shown in Figure~\ref{fig:evo285}. On the other hand, the model with  $\menv \sim 10^{-1} \msun$ does not change its luminosity or surface temperature.

\begin{figure*}[ht]
\begin{center}
  \includegraphics[width=0.9\textwidth]{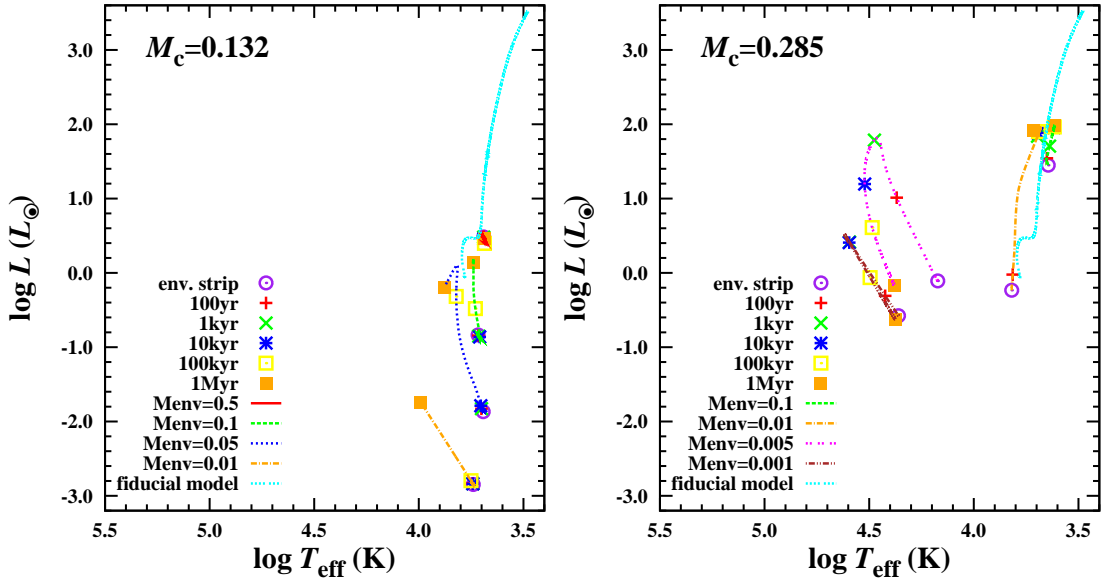}
 \end{center}
  \caption{
  Location and evolution of models with  $\mcore=0.132~\msun$ (left) and $\mcore=0.285~\msun$ (right) after the envelope stripping. The open circles denote models immediately after the envelope stripping, each of which represents a different value of $\menv(\msun)$ along the sequence from the red-giant to white dwarf branch. The lines starting from the open circles show the evolutionary tracks after the envelope stripping. The filled symbols denote the locations at $10^n$ yr ($n=2,3,4,5,6$) as indicated in the panels.
  }
  \label{fig:evohrd}
 \end{figure*}

The bumps of $\teff$ and $\log L$ in $10^{-3}~\msun<\menv<10^{-2}~\msun$ are a consequence of the thermal evolution of  model stars between the red-giant and white dwarf branch on the H-R diagram (See. Fig.~\ref{fig:evohrd}). 
All the models decrease their luminosities immediately after the envelope stripping because the timescale of the  stripping is much shorter than the thermal timescale \citep{Langer2000,Podsiadlowski2003}. Then they recover  their luminosities on the thermal timescales depending on the masses of the envelopes and cores if the hydrogen burning still continues. The envelope remains convective if its mass is larger than $10^{-3} \msun$ -  $10^{-2}~ \msun$ depending on the core mass. Thus such models quickly recover the luminosities within 100 yr and return to the red-giant branch. For models with $\mcore=0.4 ~\msun$ in which the hydrogen burning is extinguished, the envelopes become already radiative and reside on the white dwarf branch. On the other hand, models with less  massive cores have radiative envelopes even if hydrogen is still burning. Then the model with $\mcore=0.285~\msun$, for example, is fainter for less massive envelopes  ($\menv<10^{-2} ~\msun$)  partly due to lower energy generation rates and partly due to lower efficiency of the convective transport of energy.   Models with the same core mass but with thinner envelopes become brighter as long as hydrogen is burning because the radiation can more quickly transport the energy.
If the envelopes are thin enough and the hydrogen burning is extinguished, models already reside on the white dwarf branch and no longer evolve for next $100 \hyp 1,000$ yr because the thermal timescales in the envelopes are very short.
Thus the critical values of $M_{\rm env}$ or $M_{\rm hyd}$  that divide the evolution into  the two branches are larger  for a smaller core mass, which is, of course, a consequence of  $M_{\rm env, es}$ at which the energy generation rate dramatically drops (Fig. \ref{fig:nuc}).

Once the hydrogen burning is extinguished,  model stars move to the white dwarf branch on the H-R diagram as shown in Figure~\ref{fig:100HR}. The points with the same symbols represent models at $100$ yr with different values of $\menv$ starting from the same core mass.
Apparently, model stars follow the evolution from the red-giant branch to the white dwarf with strong mass loss. Thus the evolutionary path on the H-R diagram may also enable us to estimate the mass of the helium core of the companion star at the instance of the SN, which will help understand characteristics of the binary evolution.

\begin{figure*}[ht]
 \begin{center}
  \includegraphics[width=0.9\textwidth]{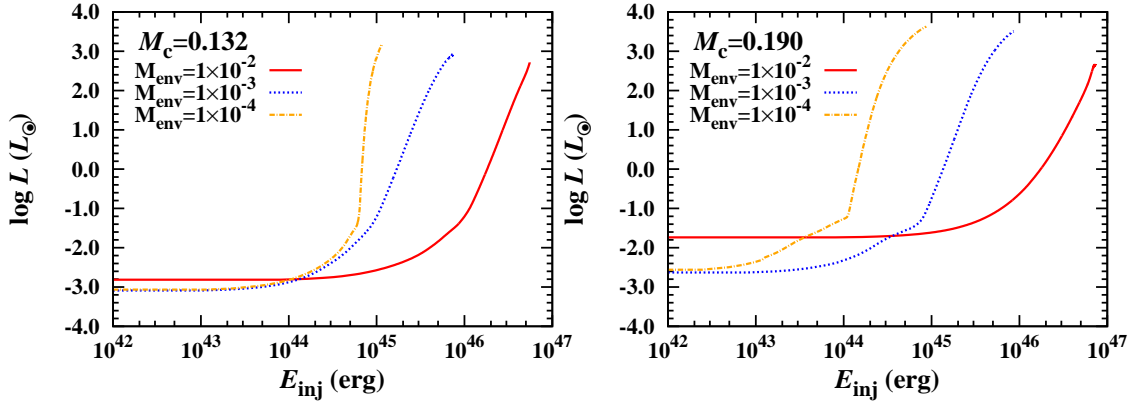}
   \end{center}
 \caption{
   Temporal change of the luminosity during the energy injection for models with $\mcore = 0.132~ \msun$ (left) and $\mcore= 0.190~ \msun$ (right) . Energy is injected  at a constant rate into the residual envelopes of the envelope-stripped companions within 1 year.
  Individual lines represent models with different values of $\menv$ indicated in each panel.
  }
  \label{fig:LEbind}
\end{figure*}
\begin{table}[ht]
\begin{center}
\begin{tabular}{c l l} \hline \hline
$\mcore$ ($\msun$) & 0.132 & 0.190 \\ \hline
 Case 1 & $5.5\times10^{46}$ & $7.4\times10^{46}$ \\
 Case 2 & $2.8\times10^{46}$ & $3.7\times10^{46}$ \\
 Case 3 & $5.5\times10^{45}$ & $7.4\times10^{45}$ \\
 Case 4 & $2.8\times10^{45}$ & $3.7\times10^{45}$ \\ \hline
\end{tabular}
\end{center}
\caption{Parameters for the injected energy (in erg) for two helium core masses. All the models in this table have the same envelope mass of $\menv=\pow{1}{-2} \msun$.}
\label{tab:model} 
\end{table}
\begin{figure*}[ht]
 \begin{center}
  \includegraphics[width=0.9\textwidth]{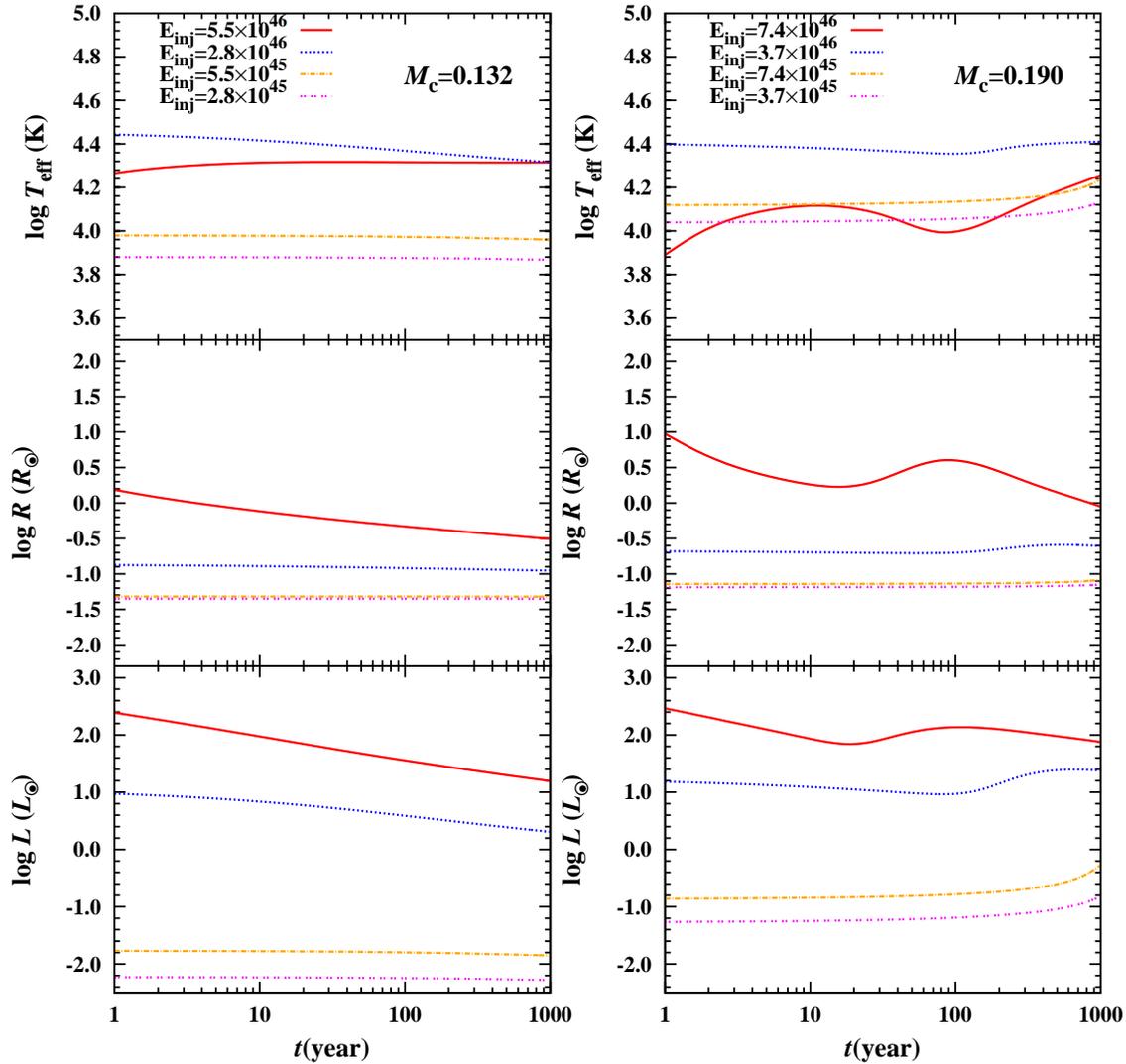}
    \end{center}
 \caption{
  Time evolution  of effective temperature (top), radius (middle) , and luminosity (bottom) for models in Table~\ref{tab:model} with the same envelope mass of $\pow{1}{-2} \msun$.
  The injected energy and the helium core mass are indicated in the top panels.
  }
  \label{fig:evoinj}
\end{figure*}
\begin{figure*}[ht]
  \begin{center}
 \includegraphics[width=0.9\textwidth]{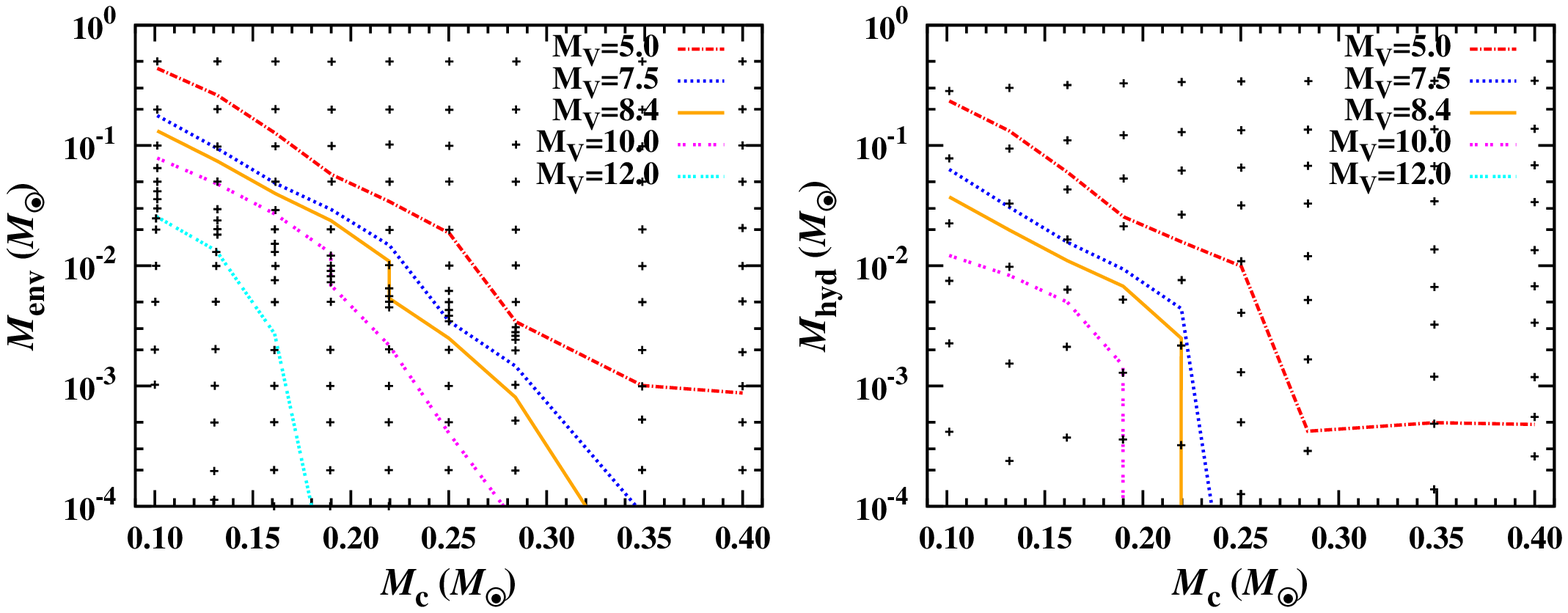}
     \end{center}
 \caption{
Contour plots of the absolute V magnitudes of models at $400$ yr after the supernova explosion as  functions of the envelope mass and  the helium core mass (left). The right panel shows the same but as a function of the hydrogen mass in the envelope and the helium core mass.
  No energy is injected into the envelope.
   The  crosses show the locations of computed models.
  }
  \label{fig:contour}
\end{figure*}

\subsection{Models with energy injection}

We  examine  six models  with $\mcore=0.132$ and $0.190 \msun$, each of which has $\menv = \pow{1}{-2}$, $\pow{1}{-3}$, or $\pow{1}{-4} \msun$
%$M_{\rm env}=1\times10^{-2},1\times10^{-3},\,{\rm and}\,1\times10^{-4}{\,M_\odot}$
to see effects of  heating by SN ejecta on the thermal evolution of the envelope.
We have chosen models fainter than the detection limit of  SNR 0509-67.5 (described later in Figure \ref{fig:contour}) without energy injection.
The resultant evolution of luminosity during the energy  injection of  the six models are shown in Figure \ref{fig:LEbind}.

We follow the evolution after the energy injection for $\sim 1,000$ yr. 
To exemplify the subsequent evolution, we pick up four models from each of those with the two core masses. These models have the same envelope mass of $\menv=1\times10^{-2}~\msun$ but with different amounts of injected energy listed in Table~\ref{tab:model}.
The total injected energy $E_{\rm inj}$ for each $\mcore$ amounts to approximately one fourth or one third of the total gravitational binding energy of the envelope.
%Models have a dimension of a red-giant or a white dwarf.
The evolution of  representative models are shown in Figure \ref{fig:evoinj}.

Models in Case 1 once expand the envelopes in response to the energy injection, but soon the envelopes shrink on the Kelvin-Helmholtz contraction timescale,
%added equation
\begin{eqnarray}
t_{\rm KH}&\sim& \frac{GM_{\rm c}M_{\rm env}}{RL}\nonumber \\
&=&60\,{\rm yr}\left(\frac{M_{\rm c}}{0.190\msun}\right)\left(\frac{M_{\rm env}}{0.01\msun}\right) \left(\frac{10R_\odot}{R}\right) \left(\frac{100L_\odot}{L}\right).\nonumber
\end{eqnarray}
In particular, the model with $\mcore = 0.190 \msun$ evolves back to a giant at first and then reduces its radius by a factor of three in 60 yr ($t_{\rm KH}$). On the other hand, models with less injected energies (other than Case 1) remain on the white dwarf branch. The energy injection from SN ejecta delays the thermal adjustment of the remaining envelope.  This also holds  for other envelope masses  such as $M_{\rm env}=10^{-3}$ and $10^{-4} \msun$.    

%\subsubsection{Lumionsity after energy injection}

 As shown in Figure~\ref{fig:LEbind}, the  luminosity suddenly increases when the amount of injected energy exceeds a certain critical value  for each model.  The critical  energy  is roughly proportional to the mass of the envelope.
This sudden increase in the luminosity can be ascribed to a change of the thermal structure of the envelope from a radiative to a convective envelope.
The model shifts to a giant once a sufficient amount of energy is  injected into the envelope.
As is shown in Figure \ref{fig:evoinj}, the luminosity of the giant monotonically declines immediately after the injection.
On the other hand, models with lower $E_{\rm inj}$ remain on the white dwarf branch.
These models keep their luminosities at the end of energy injection for several hundred years as shown in Figure \ref{fig:evoinj}.

\section{Comparison with SNR 0509-67.5}
We revisit the possibility of the existence of a surviving companion for SNR 0509-67.5, once excluded by
\citet{Schaefer2012}. 
They argued that a possible ex-companion should be  fainter than $M_{\rm V}=8.4$  at the age of $\sim400$ yr and that no such a faint ex-companion has been predicted from the SD scenario in the literature. As discussed above, the thermal evolution of the residual envelope is important for this timescale. 
We follow the evolution of  surviving red-giant companions  after the impact of SN ejecta in different circumstances parametrized by three quantities $\mcore$, $\menv$, and $E_{\rm inj}$.
By comparing our results with this observational limit, we put constraints on the SD scenario.
We calculate $M_{\rm V}$ at 400 yr after the impact as functions of these three parameters.
To obtain $M_{\rm V}$, we use a trilinear interpolation of the bolometric correction of the color transformation tables of \citet{1998A&AS..130...65L} on a grid composed of the surface gravity $\log g$, the effective temperature $T_{\rm eff}$, and [Fe/H] . We adopt the extinction parameter $R_{\rm V} = 3.1$ and the reddening parameter $E({\rm B} - {\rm V}) = 0.075$.

%\subsection{Constraints on the core mass and the envelope mass}

Figure \ref{fig:contour} shows the contour plots of $M_{\rm V}$ at 400 yr after the impact as functions of $\mcore$ and $\menv$ without energy  injection , i.e., $E_{\rm inj}=0$.
For models with $\mcore \gtrsim 0.3~ \msun$, the mass of the envelope to satisfy the observational limit for SNR 0509-67.5 needs to be smaller than $\sim 10^{-3} \msun$.
         This value corresponds to $\approx 0.2$ \% of the initial envelope mass of $1 ~\msun$ star and is comparable to the theoretical limit for the remaining mass of giants according to \citet{Marietta2000}.
Therefore it may be unlikely that an evolved red-giant with $\mcore \gtrsim 0.3~ \msun$ is a candidate for the surviving companion of SNR 0509-67.5.
For less evolved giants with $\mcore < 0.25~ \msun$, we have a plenty of the parameter space to fulfill the condition that the companion star is not detectable with the current observational upper limit, if we ignore the energy injected into the envelope. 

%\subsection{Constraints on the amount of the injected energy}

In the following, we discuss the detectability of ex-companions with core masses of $\mcore = 0.132$ and $0.190~ \msun$ taking account of the energy injected by SN ejecta. Other models with more massive cores ($\mcore>0.28~\msun$) are already too bright to satisfy the observational limit for the ex-companion in SNR 0509-67.5.

As shown in Figures \ref{fig:LEbind} and \ref{fig:evoinj}, the maximum injected energy consistent with an ex-companion in this SNR is estimated by the luminosity immediately after the energy injection by ignoring the change of the luminosity over $\approx 400$ yr after the explosion.
Table \ref{tab:injcrit} shows the critical values of the  injected energy $E_{\rm inj,\,crit}$  to satisfy the observational limit.
\begin{table}[t]
\begin{center}
\begin{tabular}{l l l} \hline \hline
$M_{\rm c}({\,M_\odot})$&$M_{\rm env}({\,M_\odot})$&$E_{\rm inj,crit}({\rm erg})$ \\ \hline
0.132&$1\times10^{-2}$&$1.1\times10^{46}$ \\
0.132&$1\times10^{-3}$&$1.1\times10^{45}$ \\
0.132&$1\times10^{-4}$&$6.4\times10^{44}$ \\
0.190&$1\times10^{-2}$&$3.9\times10^{45}$ \\
0.190&$1\times10^{-3}$&$1.0\times10^{45}$ \\
0.190&$1\times10^{-4}$&$1.2\times10^{44}$ \\ \hline
\end{tabular}
\end{center}
\caption{Critical injected energy  (in erg) for  models to have a brightness of $M_{\rm V}=8.4$ at 400 yr.}
\label{tab:injcrit} 
\end{table}

The values in Table~\ref{tab:injcrit} correspond to $\sim 1 \hyp 10$ \% of the energy that a companion star filled in the Roche lobe may receive from a SN.
Suppose that a $1 \msun$ star fills  its Roche lobe.   The star receives the kinetic energy brought by the ejecta passing through the Roche lobe. If the total energy of the SN is $10^{51}$ erg, the  energy injected into the companion is estimated to be  $\sim3\times10^{49}$ erg  by considering the geometrical cross section of the red-giant using the fomula in \citet{Eggleton1983}.
If the energy is equally distributed per unit mass  in the entire envelope, the remaining envelope  should retain $\sim3\times10^{45}$-$3\times10^{47}$ erg for $\menv=10^{-4}$-$10^{-2}\,\msun$.
Since the distribution of the energy injected into the envelope is not well understood and not necessarily homogeneous, this is only a rough estimate.
Although detailed numerical simulations are desirable to see whether such energies can be injected into tightly bound residual envelopes very close to the core edge, there are apparently some parameters  for which a surviving companion becomes too faint to be detected as a result of the envelope stripping.

\section{Conclusions}

We consider the evolution of surviving red-giant companions strongly affected by SN explosions.
The parameters we investigated are  the mass $\mcore$  of the helium core of a red-giant, the residual envelope mass $\menv$ after  the envelope stripping process, and the amount of the  energy $E_{\rm inj}$ injected into the envelope  by the SN ejecta.  

We found that the brightness of a surviving companion after the SN is mainly determined by the mass of the helium core and the residual envelope mass. Companions suddenly become  faint  during the stripping when the envelope mass is reduced down to $10^{-2}~\msun$-$10^{-4}~\msun$ depending on $\mcore$.
The critical value of the envelope mass (or hydrogen mass) decreases with increasing core mass.
After the  stripping, the envelope approaches a thermal equilibrium state on  the Kelvin-Helmholtz timescale, which results in the increase of the luminosity. Once the thermal equilibrium is achieved, the companion settles on the evolutionary track of mass losing red-giants.
On the other hand, a companion in the white dwarf branch also has a chance to evolve into a red-giant by energy injection, but the red-giant gradually shrinks on the timescale of $\sim1,000$ yr.

The results of our models are applied to the case  of SNR0509-67.5 whose detection limit to the ex-companion star is ${\rm M_{V}} = 8.4$ at $400$ yr \citep{Schaefer2012}.
 We find that this limit excludes progenitors with $\mcore \gtrsim 0.3~\msun$ due to very small ($\menv < 10^{-4}~\msun$) envelope mass allowed for surviving companions.
For $\mcore = 0.19~\msun$ models, ex-companions should have the envelope mass of $\menv<0.02~\msun$ ($M_{\rm hyd}<0.01~\msun$). 
For even smaller core mass of  $\mcore = 0.10 \msun$, the surviving stars can retain $\menv<0.1~\msun$ ($M_{\rm hyd}<0.03~\msun$) .

We estimated the maximum  energy injected into the envelope by the SN ejecta, to account for no detection of the ex-companion star for SNR 0509-67.5 and found that there exists a range of parameters for the ex-companion star even with some injected energy. If  less than 0.1\% of the kinetic energy of a part of the SN ejecta colliding with the companion is injected into the envelope, then the companion of SNR 0509-67.5 could be faint enough.

Detailed simulations of the propagation of a SN blast wave in the envelope will provide us with more stringent constraints on the parameter range to be consistent with the observations of  SNR 0509-67.5 under the SD scenario.

\bibliographystyle{apj}
\bibliography{Supernova,Whitedwarf,Others}

\end{document}